# A novel framework for fully-automated co-registration of intravascular ultrasound and optical coherence tomography imaging data

Xingwei He, MD[1,2,#]; Kit Mills Bransby, PhD[3,#]; Ahmet Emir Ulutas, MD[4]; Thamil Kumaran, PhD[4]; Nathan Angelo Lecaros Yap, MBChB[1]; Gonul Zeren, MD[4]; Hesong Zeng, MD, PhD[2], Yaojun Zhang, MD, PhD [5]; Andreas Baumbach, MD, PhD[1,4]; James Moon, PhD[6]; Anthony Mathur, PhD[4]; Jouke Dijkstra, PhD[7]; Qianni Zhang, PhD[3]Lorenz Raber, MD, PhD[8]; Christos V Bourantas, MD, PhD[1,4,*]

[1] Department of Cardiology, Barts Heart Centre, Barts Health NHS Trust, London, UK

[2] Division of Cardiology, Department of Internal Medicine, Tongji Hospital, Tongji Medical College, Huazhong University of Science and Technology, Wuhan, China

[3] School of Electronic Engineering and Computer Science, Queen Mary, University of London, London, United Kingdom

[4] Centre for Cardiovascular Medicine and Devices, William Harvey Research Institute, Queen Mary University London, UK

[5] Department of Cardiology, Xuzhou Third People's Hospital, Xuzhou, China

[6] Institute of Cardiovascular Sciences, University College London, London, UK

[7] Division of Image Processing, Department of Radiology, Leiden University Medical Center, Leiden, The Netherlands

[8] Department of Cardiology, Bern University Hospital, University of Bern, Bern, Switzerland

[#] Equal Contributions

**preprint**

**Short title:**

**Word count:**

***Address for correspondence**

Christos V Bourantas MD PhD

Consultant Cardiologist, Barts Heart Centre

Professor of Cardiology, Queen Mary University of London

Barts Heart Centre, West Smithfield, London EC1A 7BE

E-mail: c.bourantas@gmail.com

Phone: +44 20 7377 7000

Fax: +44 20 7791 9670

**Conflicts of Interests:** All authors have no conflicts of interest to declare.

**Funding:** This study is jointly funded by the British Heart Foundation; AB AM, JCM and CVB are funded by Barts NIHR Biomedical Research Centre, London, UK.

**Abstract**

**Aims:** To develop a deep-learning (DL) framework that will allow fully automated longitudinal and circumferential co-registration of intravascular ultrasound (IVUS) and optical coherence tomography (OCT) images.

**Methods and results:** Data from 230 patients (714 vessels) with acute coronary syndrome that underwent near-infrared spectroscopy (NIRS)-IVUS and OCT imaging in their non-culprit vessels were included in the present analysis. The lumen borders annotated by expert analysts in 61,655 NIRS-IVUS and 62,334 OCT frames, the side branches identified in 10,000 NIRS-IVUS frames and 10,000 OCT frames, and the calcific tissue annotations in 10,000 NIRS-IVUS frames and 10,000 OCT frames, were used to train DL solutions for the automated extraction of these features. The trained DL solutions were used to process NIRS-IVUS and OCT images and their output was used by a dynamic time warping algorithm to co-register longitudinally the NIRS-IVUS and OCT images, while the circumferential registration of the IVUS and OCT was optimized through a rotation cost matrix and dynamic programming. On a test set of 77 vessels from 22 patients, the DL method showed high concordance with the expert analysts for the longitudinal and circumferential co-registration of the two imaging sets (concordance correlation coefficient >0.99 for the longitudinal and >0.90 for the circumferential co-registration). The Williams Index was 0.96 for the longitudinal alignment and 0.97 for circumferential co-registration, indicating a comparable performance of the proposed framework to the analysts. The time needed for the DL pipeline to process imaging data from a vessel was <90sec.

**Conclusion:** The fully automated, DL-based framework introduced in this study for the co-registration of IVUS and OCT is fast and provides estimations that compare favorably to the expert analysts. These features renders it useful in research in the analysis of large-scale data collected in studies that incorporate multimodality imaging to characterize plaque composition.

**Keywords**: Deep learning; Intravascular imaging; Co-registration; Coronary artery disease

# Graphical Abstract

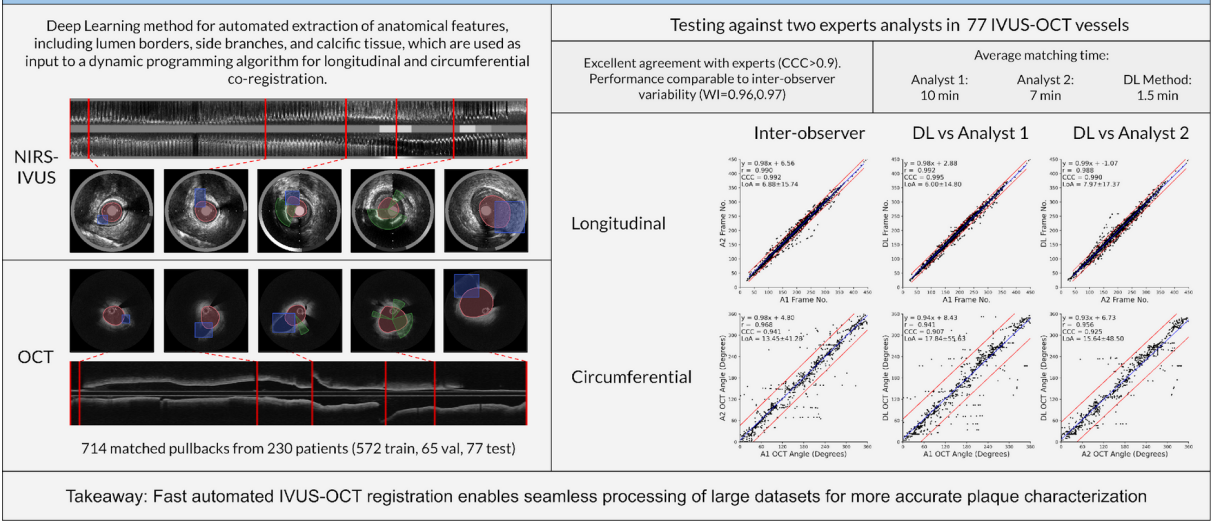

Takeaway: Fast automated IVUS-OCT registration enables seamless processing of large datasets for more accurate plaque characterization

# 1. Introduction

Intravascular imaging was introduced to in vivo assess plaque pathology and identify lesions that are likely to progress and cause events (1, 2). Prospective studies using intravascular ultrasound (IVUS), optical coherence tomography (OCT) or near infrared spectroscopy (NIRS)-IVUS have shown that the existing invasive imaging techniques are able to detect high-risk plaques with however a limited positive predictive value fact that has been attributed to the inherited limitations of these modalities to assess plaque characteristics (3-7). This has also been confirmed in histology studies which underscored the weakness but also the complementary strengths of the existing imaging techniques to assess plaque morphology. Today it is widely acknowledged that IVUS is the ideal modality for quantifying the plaque burden and detecting calcific tissue but it is unable to detect the necrotic core tissue; conversely NIRS can provide reliable detection of necrotic cores but it cannot give depth information about their distribution, while OCT with its high image resolution is the best modality for measuring the cap thickness over necrotic cores and detecting plaque micro-characteristics associated with increased vulnerability but it has limited penetration depth and thus it does not allow visualization of the entire plaque in heavily diseased segments (8, 9). To overcome these limitations hybrid imaging catheters have been introduced that combine different imaging techniques in a single probe and appear superior to standalone imaging in assessing plaque composition; however the clinical applications of these approaches are limited and today there is no imaging system that combines all the three approaches (IVUS, OCT and NIRS) in a single probe (10).

Therefore, there is a trend over the recent years to perform multimodality intravascular imaging in studies that aim to thoroughly assess plaque pathology or to examine the implications of novel pharmacotherapies on atherosclerotic disease progression (11-15). The analysis of the collected IVUS and OCT data in these studies however is performed separately despite the fact that it has been acknowledged that the co-registration of these data and the simultaneous display of the information provided by the two modalities in hybrid image is likely to allow more accurate characterization of plaque morphology (16). This is not feasible however, in research workflows as the matching of the IVUS and OCT images requires expertise and is a laborious process. To overcome this limitation semi-automated methodologies have been developed that take advantage of anatomical landmarks detected by expert analysts in IVUS and OCT to identify corresponding images in the two datasets and estimate their circumferential orientation. These approaches, however, have not been extensively validated, or they require preprocessing and segmentation of the imaging datasets and appear inferior to expert analysts in identifying matching (17-20). To overcome these limitations, we present in this study a fully automated deep learning (DL)-based framework for feature extraction and co-registration of IVUS and OCT imaging data.

## 2. Methods

*2.1 Study population*

We retrospectively analyze data from 230 patients (714 vessels) collected in the PACMAN-AMI trial (NCT03067844), an investigator-initiated, multi-center, randomized, double-blind clinical trial conducted at 9 centers in 4 European countries that aimed to investigate the effect of intensive lipid-lowering therapy with alirocumab, added to high-intensity statin therapy on plaque characteristics in patients admitted with an acute myocardial infarction (AMI) (12, 21). The recruited patients had successful percutaneous coronary intervention (PCI) of the culprit lesion and angiographic evidence of coronary artery disease without a significant obstruction (diameter stenosis >20% and <50% by visual estimate) in the proximal segment of the non-culprit vessels and high low density lipoprotein measured before intervention (≥125 mg/dL for patients that were not receiving a stable statin dose for at least 4 weeks or ≥70 mg/dL for those that were on a stable statin dose for at least 4 weeks). All patients underwent NIRS-IVUS and OCT imaging in the non-culprit vessels post PCI and were then randomized at 1:1 ratio to high intensity statin therapy and subcutaneous alirocumab therapy (150 mg biweekly) or statin monotherapy plus placebo. Treatment was given for 52 weeks and then the patients were invited for repeat angiography, NIRS-IVUS and OCT imaging of the non-culprit vessels. The study protocol complied with the Declaration of Helsinki and was approved by the local research ethics committee. All patients provided informed consent for enrollment in the institutional database for potential future investigations.

*2.2 NIRS-IVUS and OCT images data acquisition*

NIRS-IVUS and OCT imaging were performed in the proximal segments of the non-culprit vessels for a length of at least 50mm using the 40 MHz INSIGHT TVC-C195-22 or the 50 MHz INSIGHT TVC.C195-32 catheter (Infraredx, Burlington, MA, USA). After intracoronary administration of 100-200μg of nitroglycerine the catheter was advanced to the distal vessel and then was pulled back by an automated pull-back device at a speed of 0.5mm/s. In each patient the same system was used to assess the non-culprit vessels at baseline and follow-up.

OCT imaging was performed in the segments assessed by NIRS-IVUS using the ILUMIEN OPTIS (Abbott Vascular Santa Clara, CA, USA) system. After administration of intracoronary nitroglycerin, the OCT catheter was advanced to the distal vessel and pulled-back during contrast injection at a speed of 36mm/s. For optimal image quality contrast injection was performed using an automated ACIST CVi Contrast Delivery System with an injection rate of ≥5.0ml/s for the left coronary system and ≥4.0ml/s for the right coronary artery depending on the vessel size.

*2.3 NIRS-IVUS and OCT segment extraction*

NIRS-IVUS and OCT image analysis was performed at the Coronary Imaging Corelab Queen Mary University London. An expert analysts (CVB) reviewed the intravascular imaging data acquired at baseline and follow-up and used anatomical landmarks such as the coronary ostia or the origin of side branchers seen in the angiographic datasets, the NIRS-IVUS and the OCT images to define the segment of interest (SOI) that consisted of the longest segment that was assessed by both NIRS-IVUS and OCT at the two time points. The NIRS-IVUS images portraying the SOI were then processed using a dedicated deep-learning algorithm (22), that enabled retrospective identification of the end-diastolic (ED) frames, while the OCT frames were analysed in the SOI at every 0.4mm interval. The 714 SOI identified in 230 patients were then split in 3 sets: a feature training set consisted of 187 patients (572 vessels), a validation set consisted of 21 patients (65 vessels) and a test set consisted of 22 patients (77 vessels).

**Figure 1:** Snapshot of QCU-CMS software used by the expert to co-register the NIRS-IVUS and OCT data. After the identification of the ED frames in NIRS-IVUS, showing in the longitudinal IVUS image with green bars, and the selection of OCT frames at 0.4mm interval, shown in the longitudinal OCT image with red bars, in the SOI an expert analyst reviewed the IVUS and OCT images and identify corresponding frames using anatomical landmarks (side branches, and the presence of calcification). The location of these landmarks are indicated with a green dot and a number in the longitudinal IVUS and OCT frame. These landmarks were also used to define the circumferential co-registration in the NIRS-IVUS and OCT images. The frames where the expert rotated the OCT images to circumferentially align them with IVUS are shown in the longitudinal OCT image with a blue dot.

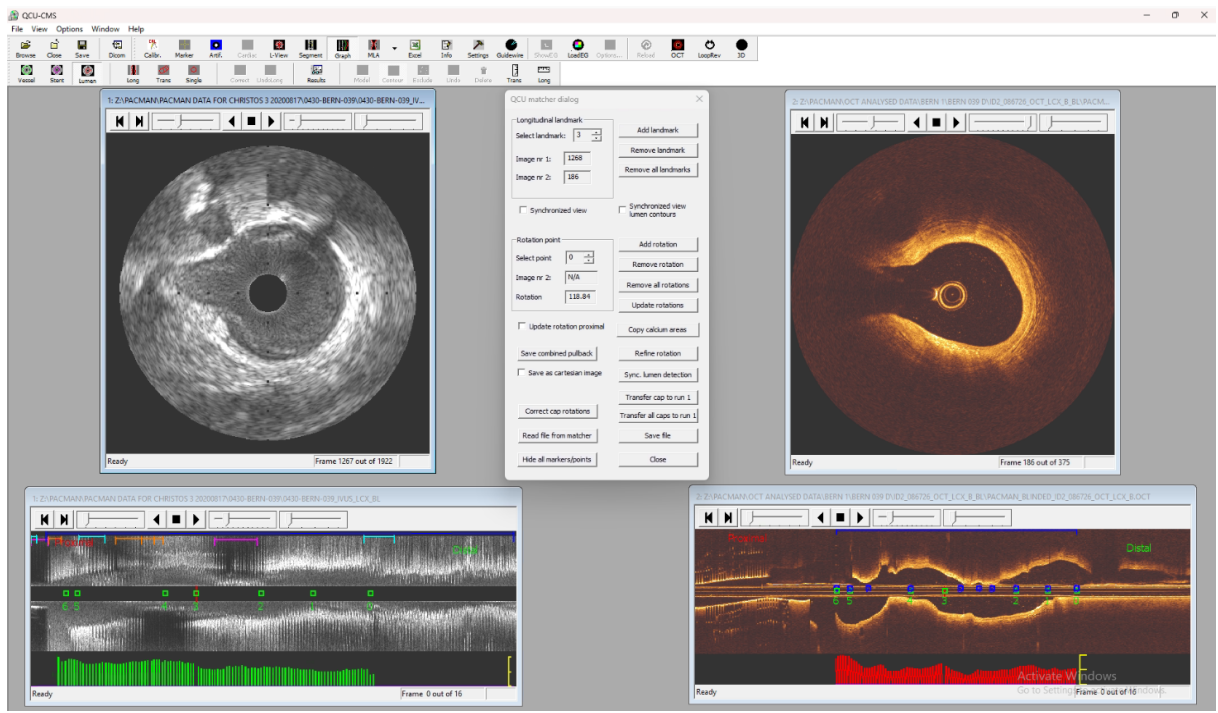

*2.4 NIRS-IVUS and OCT manual co-registration*

Data consisting of validation (21 patients, 65 vessels) and test sets (22 patients, 77 vessels) were selected for manual co-registration by expert analysts. First, an expert analyst (AU) reviewed the NIRS-IVUS and OCT images consisting of the SOI in these sets, identified corresponding frames and estimated their rotational orientation. A specially designed module of the QCU-CMS software (Version 4.69, Leiden University Medical Center, Leiden, the Netherlands) was used for simultaneous visualisation of the NIRS-IVUS and OCT pullbacks allowing the expert analyst to identify matched sections that portray anatomical landmarks such as the origin of side branches or the presence of calcific tissue that were visible in both modalities (**Figure 1**). In the test set matching was also performed by a second analyst and the first analyst performed the analysis twice so as to report the inter- and intra-observer variability. The lipid core tissue distribution was not considered in the matching process of the two modalities as there were often differences in the output of these two techniques. Linear interpolation was then applied to match the NIRS-IVUS and OCT images located between corresponding sections. In this way, each NIRS-IVUS cross section of the SOI had a corresponding OCT frame. In a final step the anatomical landmarks that were used to match NIRS-IVUS and OCT images were also used to identify the circumferential orientation of the OCT in relation to NIRS-IVUS; and then the OCT images were rotated to achieve optimal longitudinal and circumferential alignment of the two datasets. The matched NIRS-IVUS and OCT images were then used as ground truth to validate and test the effectiveness of the DL-solution for fully automated longitudinal and circumferential co-registration of NIRS-IVUS and OCT.

*2.5 NIRS-IVUS and OCT manual feature annotation*

In the feature training set, expert analysts (XH, AU, NLY) manually delineated the lumen borders in all ED NIRS-IVUS and OCT frames depicting the SOI. A subset of 10,000 frames from each modalities training set were randomly sampled. In these frames, side branches that intersected the lumen were identified and marked with a rectangular bounding box; in addition, the presence and lateral extend of calcific tissue was detected in a subset of the feature training set and represented with a label that indicated its presence or absence for every circumferential angle around the lumen centroid (**Supplementary Figure 1**).

The same process was followed in the validation and test set for all SOI frames. Manual feature annotation was performed after the co-registration of the NIRS-IVUS and OCT images so that analysts were blinded to the feature labels during the co-registration process. These labels were then used to train and validate the DL methods for the automated detection of these features (**Supplementary Table 1**).

**Figure 2**: Schematic diagram of framework used to co-register the NIRS-IVUS and OCT images. Features are extracted from NIRS-IVUS and OCT pullbacks using an ensemble of DL networks and these are then used for longitudinal registration using a DTW, and circumferential registration using a path finding algorithm.

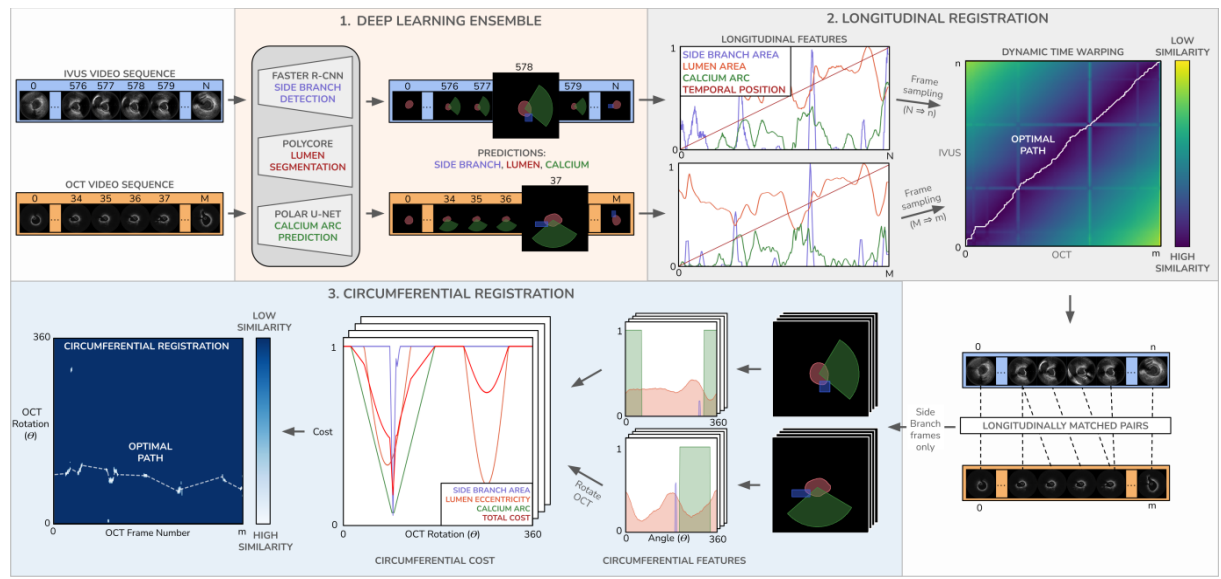

*2.6 Deep-learning methodology for NIRS-IVUS and OCT co-registration*

In contrast to the methodologies presented in the literature, the proposed methodology was designed to process NIRS-IVUS and OCT images that have not been segmented and used a multi-purpose ensemble of DL networks to extract features from the imaging data and then identify corresponding frames. A schematic diagram describing the full feature extraction and co-registration process is presented in **Figure 2**.

*2.6.1 Deep Learning Feature Extraction*

The feature extraction pipeline automatically identifies the lumen borders, the origin of the side branch, and the presence of calcific tissue in NIRS-IVUS and OCT using expert estimations (defined in Section 2.5) as supervision. Lumen border detection in NIRS-IVUS and OCT was performed using a previously validated convolutional neural network segmentation model (25). This methodology was originally trained from data acquired with the 2.4F high-resolution (35–65 MHz) Makoto™ NIRS-IVUS Imaging System (Infraredx, a Nipro Company, Burlington, MA). Because the image data used in this study were acquired with different catheters, the model was retrained for NIRS-IVUS and OCT segmentations, using the annotations provided by expert analysts in the feature training set consisted of 61,665 NIRS-IVUS and 62,334 OCT frames. Next, a fast methodology for automated detection of the origin of the side branches in NIRS-IVUS and OCT was developed taking advantage of the expert annotations in 10,000 frames in each modality (26). Histology studies indicate that both IVUS and OCT have a high-diagnostic accuracy for the detection of calcific tissue; therefore, we use this information to match NIRS-IVUS and OCT and develop a Polar-UNet network for the automated detection of the calcific tissue (8). For this network, the annotations of experts in 10,000 NIRS-IVUS frames and 10,000 OCT frames were used as input. The encoder accepts a polar view of the input image with radial and angular dimensions, while the decoder performs self-attention before classifying whether each circumferential angle around the lumen center contains calcium using expert estimations as supervision (27).

To optimize the networks, cross-entropy and dice losses were combined for lumen segmentation, a smooth L1 and cross entropy losses were used for side-branch detection, and cross-entropy loss was used for the identification of the calcific arc. Each model was trained for 200 epochs using a batch size of 64, with spatial and pixel-level augmentation. The validation set, comprising of the matched NIRS-IVUS and OCT frames of the SOI frames from 65 vessels, was used to tune network hyperparameters and select model weights based on the epoch with the lowest loss. The description of the hyperparameters and training recipe is presented in the supplementary material (**Supplementary Table 2**). The computer used to train the model contained an Nvidia A100 GPU with a 12-Core Intel Xeon CPU.

*2.5.2 Automated longitudinal and circumferential matching of NIRS-IVUS and OCT datasets*

The detected lumen borders, side branches and calcific tissue distribution in NIRS-IVUS and OCT vary longitudinally, and this information was used as input to the registration modules. To not miss the geometrical information provided in frames located between ED frames we analyzed all the NIRS-IVUS and OCT data of the SOI, and extracted four features from each modality: (1) the lumen area; (2) the side branch area, measured as the number of pixels within the predicted bounding box; both areas were normalized to the [0, 1] interval using the maximum lumen area observed in the vessel; (3) degree of calcification, quantified as the proportion of the lumen's circumference exhibiting calcified tissue; and (4) normalized frame position, representing the relative temporal location of each frame. To reduce noise and inter-frame variability, Gaussian smoothing was applied to all features along the longitudinal axis. The sequences were subsequently down-sampled by retaining the ED frames in IVUS and every second frame in OCT. Dynamic time warping (DTW) was then performed to find the optimal longitudinal alignment between the IVUS and OCT sequence (28). First, a distance matrix was computed between all sampled frames in the sequences using a feature-weighted Euclidean distance. A DTW cost matrix was then defined which describes the total minimum cumulative cost of assigning the two sequences and the optimal path of IVUS-OCT frame pairs is obtained by tracing back through the matrix. For a full description of the DTW algorithm please refer to the supplementary material.

To circumferentially register paired frames automatically, we mimic the strategy of the expert analyst by only considering pairs that contain side branch landmarks, or the presence of calcific tissue and we linearly interpolated the rotations for the remaining OCT frames. This ensures matching is optimized in high signal regions, and information-poor segments are ignored. For IVUS-OCT frame pairs predicted by DTW, we select those containing side branch landmarks or calcific tissue, and extract features that vary circumferentially such as side branch angle, calcification angle and lumen eccentricity from those pairs. A circular sampling process is defined around the lumen center, along 360° directions in 2° increments. For each direction we define the distance from lumen centroid to the lumen boundary, the side branch area, and a binary value indicating the presence of calcium in NIRS-IVUS and OCT. A rotation cost matrix R is then computed by rotating every OCT frame around each angle increment and computing the feature-weighted normalized cross-correlation with its IVUS pair. To ensure that the matching is optimized in high-signal regions, and information-poor matches are ignored, entries in R are zeroed if they do not contain side branch features. For circumferential registration, we follow an established dynamic programming method (17) where a rotational cost matrix is computed, and the lowest cost path is found by backtracking from the final to first row. The rotational movement of the path between consecutive frames is constrained by a shape regularization term (29). We tune hyperparameters on the validation set of 65 vessels, finding optimal longitudinal feature weighting of 0.3 for lumen area, 1.5 for side branch area, 0.1 for calcium degree, and 2.5 for normalized frame position. For circumferential feature weighting, 1 for side branch, 1 for calcification angle, and 0.1 for lumen eccentricity is used.

*2.7 Statistical analysis*

The distribution of continuous variables was assessed using the Kolmogorov–Smirnov test; a non-normal distribution was found and therefore results are presented as median (interquartile range) while categorical values are shown as absolute values and percentages. Comparisons between continuous variables were performed using the Mann–Whitney U test. Categorical variables were compared using the Chi-square test. The performance of the feature extraction tasks was evaluated on the test set. Specifically, the performance of the lumen segmentation method was evaluated by comparing the estimations of the experts and the DL method using the dice similarity coefficient (DSC) and Jaccard coefficient (JC) while the global average precision (AP) metric was used to evaluate the side branch detection method and global precision, recall and F1-score of the positive class and AP was implemented for the calcific tissue classification (30, 31).

To examine the efficacy of the developed method to accurately identify matched NIRS-IVUS and OCT frames we used the estimations of the two experts in the test set as a reference standard. Specifically, for every matched OCT frame we computed the distance in mm and the rotational orientation in degrees between the estimations of the analysts and the DL method. A Wilcoxon signed-rank test was used to determine whether there is significant difference between the differences in the estimations between experts and the differences between the experts and the proposed method. The concordance correlation coefficient (CCC) (32), the Sperman correlation coefficient (r) and the Williams Index (WI) (33) along with its 95% confidence interval were used to evaluate the agreement between the DL-method and the two experts. The WI measures whether the model's predictions are as similar to the human experts as the experts are with each other. A WI greater than 1 suggests that the model agrees with the experts more than the experts agree with one another. Statistical analyses were performed using Python. a P value<0.05 was considered a statistically significant difference.

## 3. Results

*3.1 Studied patients*

The baseline demographics of the patients included in the training, validation and test set are illustrated in **Table 1**. Overall, there were no significant differences between patient groups in the baseline demographics or the studied.

**Table 1**: Baseline demographics of studied patients and vessels

|  | Studied patients (n=230) | Training set (n=187) | Validation set (n=21) | Test set (n=22) | P-value |
|---|---|---|---|---|---|
| Age (years) | 58 (12) | 58 (13) | 54 (15) | 54 (11) | 0.652 |
| Gender (Male) | 194 (84.3%) | 157 (84.0%) | 19 (90.5%) | 18 (81.8%) | 0.696 |
| Current Smoker | 110 (47.8%) | 91 (48.7%) | 9 (42.8%) | 10 (45.5%) | 0.856 |
| Family history of CAD | 80 (34.8%) | 64 (34.2%) | 7 (33.3%) | 9 (40.1%) | 0.815 |
| Co-morbidities |  |  |  |  |  |
|   Diabetes | 21 (9.1%) | 15 (8.0%) | 4 (19.4%) | 2 (9.1%) | 0.251 |
|   Hypertension | 97 (42.2%) | 79 (42.2%) | 10 (47.6%) | 8 (36.3%) | 0.756 |
|   Previous PCI | 7 (3.0%) | 5 (2.7%) | 2 (9.5%) | 0 (0%) | 0.152 |
| Studied Vessels | 714 | 572 | 65 | 77 | 0.951 |
|   LAD/diagonal branch | 219 (30.7%) | 178 (31.1%) | 19 (29.2%) | 22 (28.6%) |  |
|   LCX/intermediate/obtuse marginal | 264 (37.0%) | 213 (37.2%) | 23 (35.4%) | 28 (36.4%) |  |
|   RCA | 231 (32.4%) | 181 (31.6%) | 23 (35.4%) | 27 (35.1%) |  |
| Matched NIRS-IVUS and OCT frames | 77,627 | 61,665 | 6,863 | 9,099 | 0.999 |
|   LAD/diagonal branch | 25,789 (33.2%) | 20,807 (33.7%) | 2,243 (32.7%) | 2,739 (30.1%) |  |
|   LCX /intermediate/obtuse marginal | 25,333 (32.6%) | 20,379 (33.0%) | 2,123 (30.9%) | 2,831 (31.1%) |  |
|   RCA | 26,505 (34.1%) | 20,479 (33.2%) | 2,497 (36.4%) | 3,529 (38.8%) |  |

**Table footnote:** CAD, coronary artery disease; LAD, left anterior descending coronary artery; LCx, left circumflex; NIRS-IVUS, near infrared spectroscopy intravascular ultrasound; OCT, optical coherence tomography; PCI, percutaneous coronary intervention; RCA, right coronary artery.

*3.2. Efficacy of the DL methodology to detect lumen borders, side branches and calcific tissue*

The performance of the DL solutions developed to detect the lumen border, the side branches and the calcific tissue distribution is illustrated in **Table 2**. A high DSC and JC was noted between the estimations of the expert and the developed DL methodology for the detection of the lumen borders that were similar in NIRS-IVUS and OCT, while the AP were numerically higher in OCT indicating that the DL solution introduced for the detection of the side branches performed better in the OCT data. In NIRS-IVUS images, the AP was 0.58 suggesting that the DL solution had also good performance in identifying the origin of the side branches in that dataset. Finally, the precision, recall and F1 score were close to 0.9 in NIRS-IVUS and numerically smaller in OCT indicating that the methodology developed for the detection of the calcific borders was very accurate in NIRS-IVUS and that calcific tissue detection is more challenging in OCT especially when the tissue is deeply embedded.

**Table 2:** Performance metrics of the DL methods developed for the detection of the lumen borders, the side branch location and the calcific tissue in NIRS-IVUS and OCT frames.

|  | Lumen segmentation | | Side branch detection | Calcium arc classification | | | |
| --- | --- | --- | --- | --- | --- | --- | --- |
|  | DSC | JC | AP | Precision | Recall | F1 | AP |
| NIRS-IVUS | 0.96 (0.02) | 0.95 (0.04) | 0.58 | 0.87 | 0.89 | 0.88 | 0.95 |
| OCT | 0.98 (0.01) | 0.97 (0.01) | 0.74 | 0.70 | 0.66 | 0.68 | 0.73 |

*3.3 Intra- and inter-observer variability*

The intra- and inter-observer variability of the expert analysts for the longitudinal and circumferential registration of the NIRS-IVUS and OCT frames in the test set is shown in **Table 3**. A median difference of 4.5 frames was reported between experts for the longitudinal registration, and of $9.8°$ for the circumferential registration. The intra-observer variability for these metrics was 2.3 frames and $7.2°$ respectively. A high CCC and spearman correlation were noted for the expert annotations while linear regression analysis indicated a slope close to 1 and a y-intercept close to 0 for both longitudinal and circumferential registration (**Figure 3**).

**Table 3:** Longitudinal and circumferential registration results.

|  | A1 vs A2 | A1 vs A1* | DL vs A1 | DL vs A2 | WI (95% CI) |
| --- | --- | --- | --- | --- | --- |
| OCT frame difference | 4.5 (6.4) | 2.3 (2.8) | 4.1 (3.2) | 5.3 (4.8) | 0.96 (0.94, 1.00) |
| OCT angle difference (°) | 9.8 (9.6) | 7.2 (7.6) | 9.9 (11.9) | 10.2 (8.5) | 0.97 (0.96, 0.99) |

*3.4 Comparison of the estimations of the expert analysts and the DL-methodology*

The results from the comparison of the estimations of the expert analysts and the DL method are shown in **Table 3**, and **Figure 3**, while a case example showing the estimations of the two experts and the DL method is presented in **Figure 4**. For the longitudinal matching of NIRS-IVUS and OCT, a high correlation was found between the DL-method and the estimations of the first (CCC>0.99, r>0.99) and second analyst (CCC>0.99, r>0.99). Similarly for the circumferential orientation of the OCT frames, there was a high correlation between the DL-method and the two analysts (CCC>0.90, r>0.90). No statistically significant difference was found when we compared the inter-observer variability with the differences between the DL-method and the first analyst for the longitudinal matching (p=0.395), however the differences were larger when we compared the inter-observer variability and the differences between the output of the DL method to the second analyst (p=0.001). In addition, there was no significant difference when we compared the inter-observer variability with the differences in the estimations of the DL and the first or second analyst for the circumferential orientation of the OCT frames (p=0.244, and p=0.970 respectively). The WI for the longitudinal matching of the NIRS-IVUS and OCT frames was 0.96, and for the circumferential orientation of the NIRS-IVUS and OCT the WI was 0.97 indicating that the method has a similar efficacy with the experts in identifying correspondence between NIRS-IVUS and OCT. The mean time needed to segment the NIRS-IVUS and OCT frames, identify matching in NIRS-IVUS and OCT and circumferentially co-register the corresponding frames was 84 sec using a Nvidia 1080TiGPU with a 28-core Intel i9-7940X CPU. The time taken for the first and second analyst to identify matching and circumferentially co-register the corresponding frames was 576 and 324 seconds respectively.

**Figure 3:** Linear regression analysis between the estimations of the experts (A, A") the first and second annotation of the first expert, (B, B") the estimations of the DL method and the 1st expert (C and C") and the estimations of the DL method and the second expert (D and D") for the longitudinal (top panels) and circumferential orientation (bottom panels) of the intravascular imaging data. The blue line represents the regression line and the black lines the limits of agreements (±1.96SD). The Sperman correlation coefficient (r) and CCC values are also displayed.

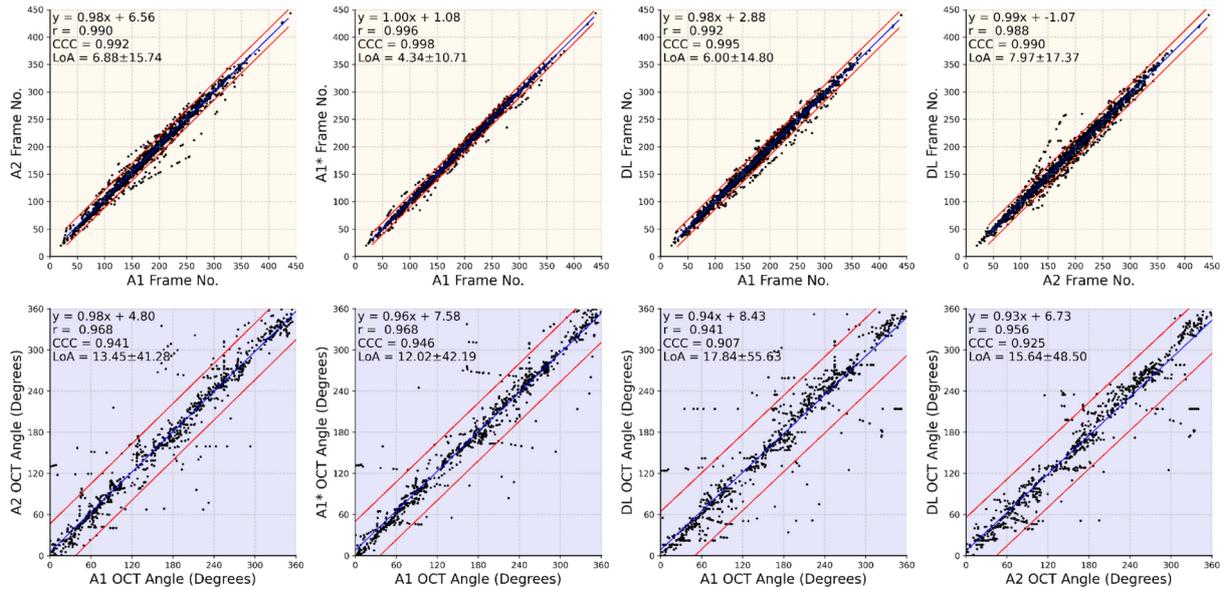

## 4. Discussion

This work introduces a fully automated framework for the longitudinal and circumferential co-registration of intravascular imaging data acquired by different invasive imaging modalities with complementary strengths. Testing against the estimations of the expert analysts in a large set of NIRS-IVUS and OCT data collected in a multimodality intravascular imaging study demonstrated that the proposed method 1) is as accurate as the experts in identifying corresponding NIRS-IVUS and OCT frames 2) it is able to identify with a high efficacy the circumferential orientation of OCT in NIRS-IVUS and 3) it is fast and fully reproducible.

Accurate assessment of plaque morphology and composition is essential in stratifying cardiovascular risk and evaluating the implications of novel pharmacotherapies on atherosclerosis disease progression (3-7, 34). Traditionally this is performed by IVUS or OCT two imaging modalities that have an established efficacy but also significant limitations in characterizing plaque morphology (8, 9). To overcome these, multimodality imaging probes have been introduced that combine different modalities with complementary strengths and appears able to enable not only more accurate evaluation of different tissue types but also assessment of plaque biology (10). Combined NIRS-IVUS, IVUS-OCT, OCT-NIRS, florescence lifetime imaging-OCT and near infrared fluorescence-OCT systems have already been introduced in clinical practice and are expected to dominate in the future in the study of atherosclerosis. However, even these systems have limitations, and are unable to provide a complete and detailed evaluation of plaque pathology. Therefore, the recent trend is to use multiple multimodality or standalone imaging probes to characterize plaque types and assess their changes over time. This strategy has allowed us to better evaluate vulnerable lesion morphology and examine the implications of emerging pharmacotherapies in plaque morphology and biology (11-15, 35). Currently, the analysis of the intravascular imaging data from these studies is performed separately for each modality as the investigators are unable to combine the collected images in a single hybrid image that will allow complete and more accurate description of plaque pathobiology. Efforts have been made over recent years to develop advanced methodologies that will be able to automatically match the intravascular imaging data acquired by different catheters (17-20). However, all of them have significant limitations as some required manual identification of frames with anatomical landmarks, others segmentation of the IVUS and OCT frames before matching, some others were unable to perform circumferential co-registration, while most of them have not been robustly validated against the estimations of expert analysts. The method of Molony et al is the most robust approach presented in the literature for the matching of IVUS and OCT frames; however this necessitates the segmentation of the IVUS and OCT images and appears to provide estimations that are inferior to the expert analysts (17). The methodology presented in this report is the first that overcomes the above limitations as it introduces a fully automated framework for fast and accurate matching of NIRS-IVUS and OCT data. It includes effective DL solutions for the detection of the lumen borders for the identification of the origin of side branches and for the presence of calcific tissue in intravascular images and takes advantage of this information to match NIRS-IVUS and OCT and estimate the rotational orientation of the OCT frames in relation to IVUS. Testing against the estimations of expert analysts in 77 vessels has shown that the proposed method is as effective as the experts in identifying corresponding frames and that it has a high performance in estimating the rotational orientation of OCT in relation to NIRS-IVUS. A significant advantage of the developed method is the fact that it is fast as it is able to process and match NIRS-IVUS and OCT images in only 1.5 minutes. These features render it useful in research; as the proposed approach can be combined with existing DL solutions for the detection of the external elastic lamina borders, the characterisation of plaque composition in NIRS-IVUS and the estimation of the thickness of the fibrous cap in OCT and combine this information in a hybrid image that will enable more reliable quantification of different plaque components and evaluation of plaque vulnerability (**Figure 5**) (19, 25, 36-38).

**Figure 4.** Comparison between method and expert analysts. Four frames (red, orange, blue, red) matched longitudinally and circumferentially in NIRS-IVUS and OCT by the first analyst (A1) are compared to the OCT frame and rotation predictions of the second analyst (A2) and the DL method. Due to interpolation, multiple OCT frames can be matched to a single NIRS-IVUS frame.

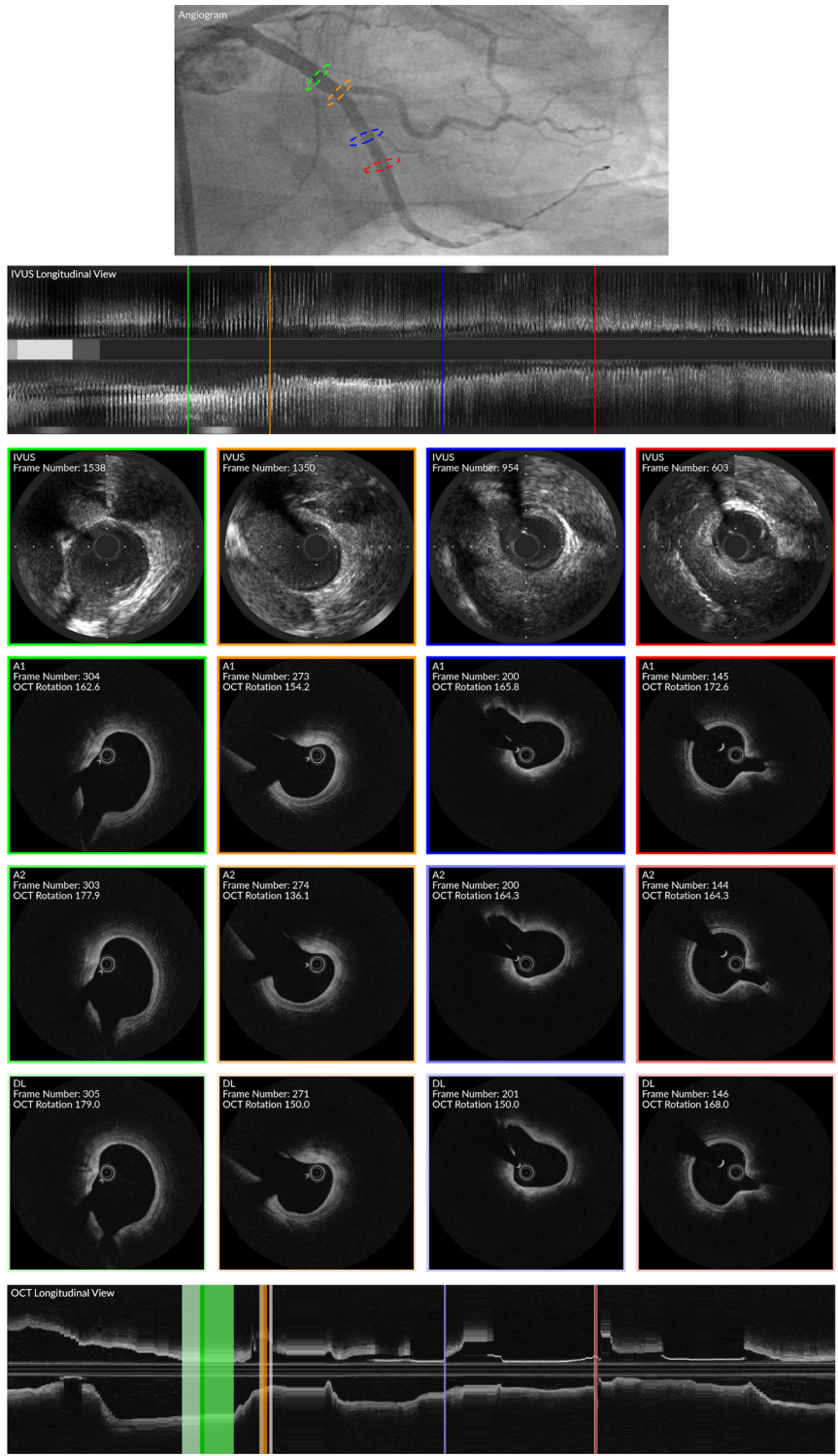

*Limitations*

Several limitations of the present analysis should be acknowledged. Firstly, the longitudinal and circumferential registration in the proposed approach is performed sequentially rather than simultaneously, therefore temporal errors are likely to propagate and increase the errors in circumferential matching. Secondly the DTW approach enforces 1-to-1 longitudinal matching between IVUS and OCT sequences, this may affect the performance of the method in segments with a small number of branches and calcific deposits. These limitations will be addressed in future work with a graph matching deep learning methodology which learns the partial assignment between sparse landmarks, rather than frames. Finally, the proposed method was trained and tested in data acquired by specific NIRS-IVUS and OCT systems therefore it is unclear whether it will be equally effective in data acquired by different catheters.

**Figure 5:** NIRS-IVUS and OCT Registration by expert. (A) NIRS-IVUS image showing lipid-rich plaque detected by near-infrared spectroscopy (NIRS), indicated by the yellow-red color scale on the outer ring. (B) IVUS-derived plaque characterization using echogenicity: green represents fibrous tissue, and red represents lipid-rich plaque. (C) OCT cross-sectional image with manually delineated fibrous cap thickness. (D) IVUS image with plaque composition analysis: region 1 (blue) corresponds to the fibrous cap. (E) Longitudinal NIRS chemogram with yellow indicating areas of high lipid content. (F) Longitudinal OCT image with numbered landmarks representing side branches. (G) Longitudinal IVUS image showing the same landmarks as in (F); identical numbers indicate corresponding side branches in IVUS and OCT views.

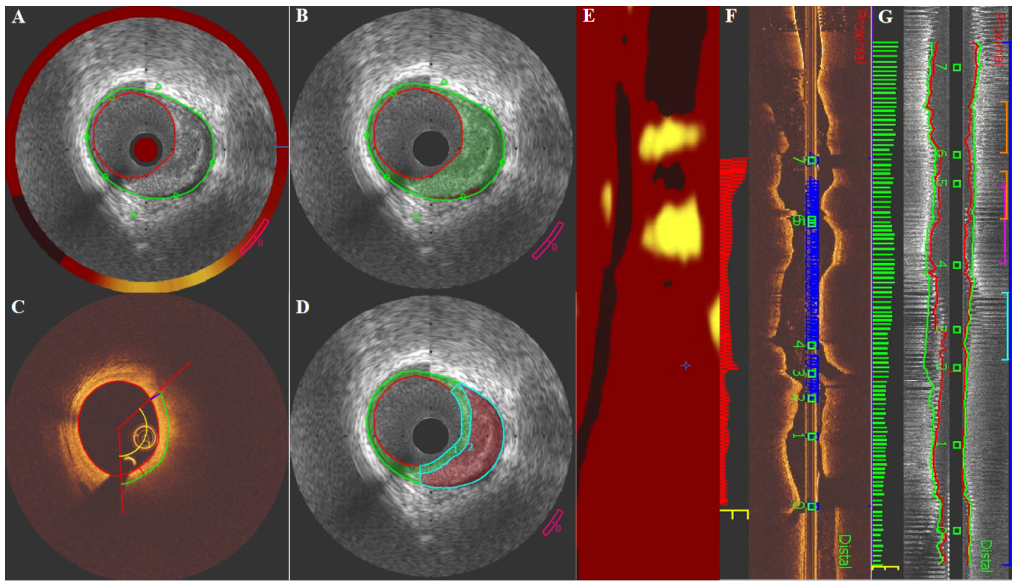

## 5. Conclusion

We introduce a fully automated framework for the longitudinal and circumferential co-registration of intravascular images acquired by an IVUS- and OCT-based systems. The proposed method is fast and provides estimations that compare favorably with the expert analysts; these unique features are expected to allow their broad use in the analysis of data collected in studies of atherosclerosis that employ multimodality intravascular imaging to assess plaque phenotypes.

**Supplementary Material**

*S1. Implementation of DTW algorithm for longitudinal registration*

DTW is performed to find the optimal temporal alignment between the NIRS-IVUS $X^t \in R^{n \times 4}$ and OCT sequence $Y^t \in R^{m \times 4}$. First, a distance matrix $D \in R^{n \times m}$ is computed between all pairs of frames in the sequence using a feature-weighted Euclidean distance as follows:

$$D_{ij} = \sqrt{\Box}$$

Where $w^t \in R^f$ is a feature-wise weighting vector, and $i \in [1,\ldots,n]$, $j \in [1,\ldots,m]$ index $X^t$ and $Y^t$ respectively. A DTW cost matrix $C \in R^{n \times m}$ is filled where each entry represents the cumulative cost to reach that point and $C_{n,m}$ is the total minimum cumulative cost of aligning the two sequences. The DTW cost matrix C is defined as:

$$C_{i,j} = D_{i,j} + \{C_{i-1,j}, C_{i,j-1}, C_{i-1,j-1}\}, C_{1,1} = D_{1,1}$$

The optimal path $P$ is a set of index pairs $(i_p, j_p)$ of length $L$ that are obtained by tracing back using the recurrence relation from $C_{n,m}$ to $C_{1,1}$:

$$P = \{(i_p, j_p)\}_{p=1}^{L}, (i_p, j_p) = arg\, arg\, \{C_{i-1,j}, C_{i,j-1}, C_{i-1,j-1}\}$$

$P$ defines the longitudinally match IVUS and OCT pairs that are used as input to the circumferential registration module.

**Supplementary Table 1:** Number of NIRS-IVUS and OCT frames included in the feature training, validation, and test set.

|  | NIRS-IVUS | | | OCT | | |
| --- | --- | --- | --- | --- | --- | --- |
|  | Training set | Validation set | Test set | Train set | Validation set | Test set |
| Lumen Segmentation | 61,665 | 6,863 | 9,099 | 62,334 | 6,484 | 8,929 |
| Side Branch Detection | 10,000 | 6,863 | 9,099 | 10,000 | 6,484 | 8,929 |
| Calcium Arc Classification | 10,000 | 6,863 | 9,099 | 10,000 | 6,484 | 8,929 |
| Co-Registration | - | 6,863 | 9,099 | - | 6,484 | 8,929 |

**Supplementary Table 2:** Training Recipes. LR=Learning Rate.

|  | Side Branch Detection | Lumen segmentation | Calcium detection |
| --- | --- | --- | --- |
| Learning Rate | 0.0001 | 0.005 | 0.005 |
| LR Scheduler | StepLR (gamma=1, stepsize=10) | Polynomial (power=0.9) | Linear |
| Batch Size | 64 | 64 | 64 |
| Epochs | 200 | 200 | 200 |
| Weight Decay | 0 | 0 | 0 |
| Optimizer | Adam | SDG | Adam |
| Loss | Cross Entropy, Smooth L1 | Cross Entropy, Dice | Cross Entropy |
| Image Resolution | 480x480 | 480x480 | 480x480 |
| Augmentation | Random Rotation (90,180,270), Brightness, Contrast | Random Rotation [-180,180], Brightness, Contrast | Random Rotation [-180,180], Brightness, Contrast |
| Confidence Threshold | 0.8 | 0.5 | 0.9 |

**Supplementary Figure 1:** Estimations of the expert analyst for the lumen, side branch and calcific tissue in paired NIRS-IVUS and OCT frames.

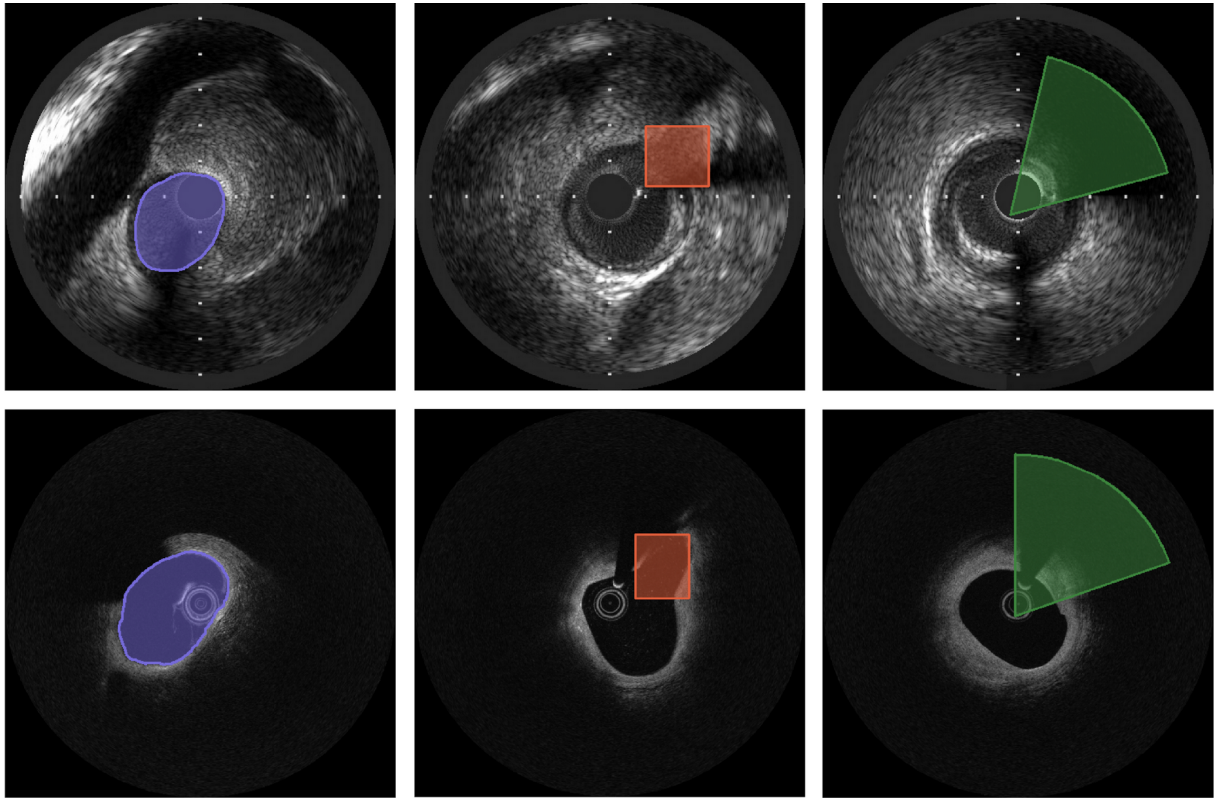